\documentclass[showpacs,twocolumn,preprintnumbers,amsmath,amssymb,aps]{revtex4}
\usepackage{epsfig}
\newcommand{\e}{{\rm e}}

\begin{document}

\author{David Biron$^*$}
\author{Elisha Moses$^*$}
\author{Itamar Borukhov$^\dagger$}
\author{S. A. Safran$^\dagger$}

\affiliation{Departments of Physics of Complex Systems$^*$ and
Materials and Interfaces$^\dagger$, Weizmann Institute of Science}

\title{Inter-filament Attractions Narrow the Length Distribution of Actin Filaments
}

\begin{abstract}
We show that the exponential length distribution that is typical of actin filaments
under physiological conditions dramatically narrows in the presence of
(i) crosslinker proteins
(ii) polyvalent counterions or
(iii) depletion mediated attractions.
A simple theoretical model shows that in equilibrium, short-range attractions
 enhance the tendency of filaments to align parallel to each other,
  eventually leading to an increase in the average filament length and a
  decrease in the relative width of the distribution of filament lengths.
\end{abstract}

\date{\today}

\pacs{xx.xx}

\maketitle

\begin{center}
{\bf Introduction}
\end{center}
The protein actin in its filamentous
form ({\em F-actin}) is a major structural component in the
cytoskeleton network, and plays an active role in maintaining the
leading edge of moving cells \cite{Lodish00}. In all these
structures, the filaments are dynamic objects that continuously
polymerize and depolymerize as part of their normal function. The
physical properties of the overall structure (mechanical strength,
viscoelastic response, etc.) depend on the properties of the
filaments that form the structure and, in particular, on their
length distribution.

\begin{figure}
\includegraphics[width=3.0in]{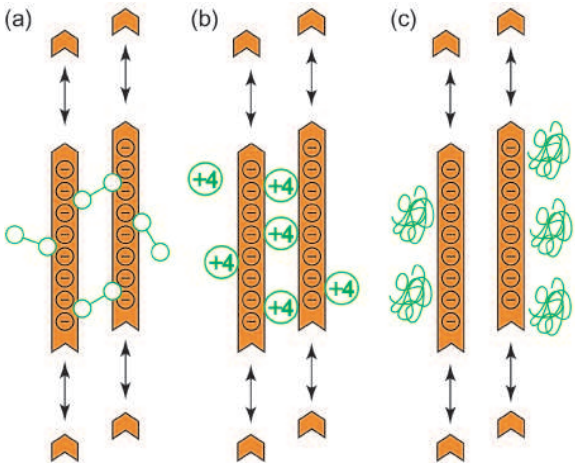}
\vspace{\baselineskip} \caption{\protect\footnotesize A schematic
view of three different attraction agents ("linkers") and their
effect on actin polymerization: (a) crosslinker proteins (b)
polyvalent counterions, and (c) inert polymers giving rise to
depletion mediated attractions.} \label{fig:schematic}
\end{figure}

\emph{In vitro} experiments on F-actin under physiological
conditions (with and without capping proteins) reveal a wide,
exponential distribution of lengths
\cite{Xu99,Littlefield98,EdelsteinKeshet98}. This can be
attributed theoretically to the fact that the underlying
(dominant) stochastic dynamics, namely monomer exchange, is an
``homogeneous zero range process'' (ZRP) \cite{Evans00} which
generates an exponential steady state distribution. At the same
time, equilibrium theories of self-assembly of polymers or
micelles predict an exponential distribution determined by the
competition of the end-cap energy and the translational entropy
\cite{SafranBook}. Indeed, nucleating and severing proteins such
as gelsolin \cite{Kreis99} are known to modify the average
filament length, $\langle l \rangle$, but not the coefficient of
variance of the length distribution
$r_\sigma\equiv\sigma_l/\langle l\rangle$, where $\sigma_l$ is the
standard deviation of the distribution.

In contrast, we have recently observed that when crosslinker
proteins are added to an F-actin solution they enhance the
preference for overlapping filament sections. As a result, the
shape of the distribution changes and $r_\sigma$ is reduced
approximately by half \cite{Biron04}. Similar behavior is also observed
in the presence of myosin aggregate \cite{Kawamura70}.

In this letter, we argue that the narrowing of the length distribution
is a general effect arising from short-range attraction between the filaments;
chemically crosslinking is not necessary to observe the reduction in $r_\sigma$.
Specifically, we show experimentally that both
depletion mediated attractions in a solution
containing inert polymers (PEG), and
electrostatic interactions induced by multivalent
counterions (spermine) result in attractions
that significantly narrow the distribution.
A simple theoretical approach demonstrates
that this effect should be observed for {\em any}
mechanism that leads to short-ranged attraction.
The attractions increase the tendency of filaments
to lie parallel to each other and to grow concurrently.
As a result, the shape of the length distribution changes
dramatically:
from a monotonously decreasing exponential distribution,
to a non-monotonous Gaussian-like distribution
with a well-defined peak at large filament lengths.
Generally speaking,
exponential distributions where the width scales as the average,
are typical of systems where entropy dominates over interaction energies;
Gaussian-like distributions, where the width is much smaller than the
average, are typical of systems where interaction energies dominate.

Conflicting experimental measurements for the values of the
elastic constants of the actin gel have been suggested to be a
result of variations in the length distribution of F-actin
\cite{Xu98}. Furthermore, cells provide a biochemical environment
crowded with macromolecules. Crowding leads to depletion mediated 
attractions that can be reproduced \emph{in vitro} by adding 
inert polymers such as polyethylene glycol (PEG) \cite{Tang04}.
We therefore suggest that the physical effects described in this 
study have a role in regulating the properties of actin gels 
in cells and in tissues.

\begin{center}
{\bf Experimental materials and methods}
\end{center}
Spermine ($4$ HCl) and PEG (MW$=6000$) were purchased from
Sigma-Aldrich Ltd. (Rehovot, Israel). Pyrene labelled actin ($10
\%$ labelled monomers) was purchased in lyophilized form from
Cytoskeleton Inc. (Denver CO, U.S.A) and polymerized at a stock
concentration of $0.15 mg/ml$ in $0.4 X$ polymerization buffer
with $0.5 mM$ ATP  (``AP-buffer'') as described in \cite{Biron04}.
G-actin aliquots were quick frozen in liquid $N_2$, and pre-spun
at $150000 g$, $4^o$C, for $2$ hours upon thawing. After
polymerization, equilibrated spermine ($100 mM$) or PEG ($4 \%
w/w$) were added to the F-actin solution from $10X$ stocks
(diluted in AP-buffer) and an equal volume of buffer was added to
control F-actin aliquots. The filaments were then incubated for at
least two hours at room temperature before performing a
depolymerization assay.

\begin{figure}
\includegraphics[width=85 mm]{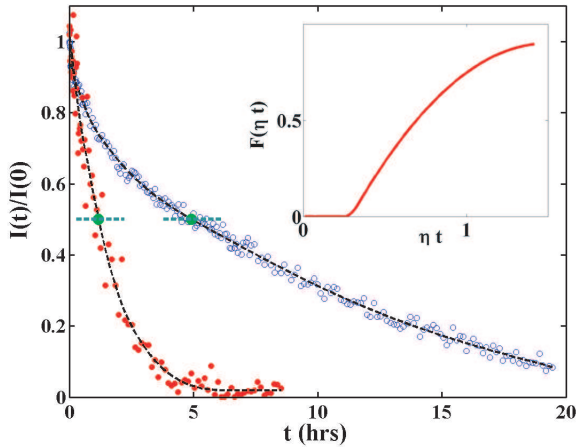}
\caption{Normalized fluorescence data from an actin
depolymerization measurement. The full circles denote F-actin
which was incubated in the presence of $100 mM$ spermine and
depolymerized by diluting the solution $10$-fold with $0.4 X$
AP-buffer. Empty circles denote a control measurement where the
F-actin solution (without spermine) was diluted with a buffer
containing $10 mM$ spermine. The dashed curves show the
non-parametric fits to the data and the straight lines are drawn
where the intensity of fluorescence reaches half of its initial
value. Inset: an example of a cumulative distribution functions 
of filament lengths which was extracted from the data. 
The depolymerization rate, $\eta$, is normalized as explained in \cite{Biron04}.}
\label{fig:exp}
\end{figure}

Depolymerization was induced by diluting the F-actin solution
$10$-fold using a cut pipette tip. F-actin aliquots incubated with
PEG (spermine) were compared to control filaments depolymerized in
AP-buffer containing $0.4 \%$ PEG ($10 mM$ spermine).
Depolymerization was monitored with a Spex FluoroLog-3
spectrofluorometer, Jobin Yvon Ltd. (London) at $20^oC$, while
stirring very gently to avoid filament breaking. An OD $1$ ND
filter was used to minimize photobleaching. Photobleaching was
measured separately and corrected for when found to be
significant. The data was non-parametrically fitted to a smooth,
monotonically decreasing and convex curve. $r_\sigma$ was then
derived from the smooth curve as described in \cite{Biron04}.

\begin{center}
{\bf Experimental results}
\end{center}
Fig.~\ref{fig:exp} presents data from an actin depolymerization
assay in the absence and in the presence of spermine ions.
We analyzed $5$ depolymerization experiments after incubation with
spermine ions, $4$ spermine control depolymerizations, $6$
depolymerization experiments after incubation with PEG and $6$ PEG
control depolymerizations (as described in Sec. II).
The values extracted for $r_\sigma$ and $\tau_{1/2}$ are
summarized in Table 1. In both cases, $r_\sigma$ was about 2-fold
smaller and $\tau_{1/2}$ was about 3-fold smaller as compared with
the control depolymerization measurements. These results are
consistent with results obtained for crosslinked filaments
\cite{Biron04}. When the length distribution is narrower the
depolymerization process is more efficient since the dominant
portion of the filament length distribution depolymerizes
concurrently, resulting in smaller values of $\tau_{1/2}$. As
explained below, narrow length distributions are a signature of
attractive inter-filament interactions, whether they result from
depletion forces \cite{Suzuki96} or from multivalent counterions
\cite{Angelini03}; physical crosslinking is {\em not} a necessity.

\begin{table}
\begin{tabular}{|c|c|c|c|}
\hline
Incubation  & Dilution  &  $ r_\sigma $ & $ \tau_{1/2} (hr)$  \\
\hline \hline
 -- &  $2 \%$ $\alpha$-actinin & $1.6 \pm 0.2$ \cite{Biron04} & $1.5 \pm 0.4$ \cite{Biron04}\\
\hline
$20 \%$ $\alpha$-actinin  & -- & $1.0 \pm 0.2$ \cite{Biron04} & $0.5 \pm 0.2$ \cite{Biron04} \\
\hline -- &  $0.4 \%$ PEG & $1.9 \pm 0.2$ & $1.5 \pm 0.4$ \\
\hline
$4 \%$ PEG  & -- & $1.1 \pm 0.2$ & $0.5 \pm 0.2$ \\
\hline
 -- & $10 mM$ spermine & $1.2 \pm 0.3$ & $5.5 \pm 1.5$ \\
\hline
 $100 mM$ spermine & -- & $0.5 \pm 0.2$ & $1.2 \pm 0.5$ \\
\hline
\end{tabular}
\caption{Results from depolymerization measurements under four
different sets of conditions. The ``incubation'' (``dilution'')
column specifies what was added to the $0.4 X$ AP-buffer during
the incubation (dilution) of the actin filaments. The values for
the coefficients of variance and the depolymerization half-times
given in each case are averaged over $4-6$ experiments for each
set of conditions and the errors are statistical.}
\end{table}

The average amplitude of the fluctuations around the smooth fit to
the depolymerization curve was similar for dilution buffer
containing 10mM spermine and not containing spermine. However, the
presence of 0.4\% PEG in the dilution buffer introduced a noise
amplitude approximately twice as large. The effect of noise is
more pronounced at the late stages of the measurement resulting in
an apparent longer tail of the distribution of lengths, which
leads to an artificially larger value of $r_\sigma$ (see
\cite{Biron04}). This is reflected in the fact that the value of
$r_\sigma$ in the well established case of the exponential length
distribution of pure F-actin was larger than unity when extracted
from dilution in the presence of PEG. It is important to note that
despite this effect, an unambiguous decrease of both $r_\sigma$ and
$\tau_{1/2}$ is preserved. This indicates that the interaction
does indeed reduce the variance of the lengths.

\begin{center}
{\bf Theory}
\end{center}
Actin polymerization is a dynamic process and most theories that
address filament length distributions
\cite{FloryBook,Oosawa,EdelsteinKeshet98} are based on kinetic
equations for the various processes involved (e.g., monomer
association, dissociation etc.). On the other hand, Flory has
already noted \cite{FloryBook} that quite often, kinetic arguments
leading to a steady state distribution can be replaced by
equilibrium arguments. In particular, the presence of
actin-severing proteins (gelsolin, cofilin, ...) ensures that the
filament distribution follows the equilibrium distribution and
does not depend on dynamic instabilities. Indeed, the exponential
distribution that is typical of F-actin is also found in
self-assembling systems where surfactant molecules form
linear micelles and equilibrium is maintained through
exchange of surfactant molecules with the solution
\cite{SafranBook}. The equilibrium approach has the advantage that
it provides a natural framework to {\em systematically} study 
the effect of inter-filament interactions.

The theoretical starting point is the following free energy (per unit volume)
of a solution consisting of concentrations, $\rho_l$, of filaments
of lengths $l=1,2,3,...$:
\begin{eqnarray}\label{Free}
    f &=& \sum_l\rho_l\left[ \ln(\rho_l v_0)-1 + \alpha l + b \right]
      \nonumber\\
      &+& \sum_{l<l'}w(l,l')\rho_l\rho_{l'} - \mu\sum_l l\rho_l
\end{eqnarray}
The logarithmic term represents the translational entropy
of the filaments where $v_0$ is a monomer volume.
Here and in the following, all energies are
in units of the thermal energy $k_BT$.
The energy of a single filament consists of
a term linear in $l$ which is due to the interaction
between neighboring monomers
and an $l$-independent term that is basically the end-cap energy, $b$, of the filaments.
The next term is the two-body interaction term where
$w(l,l')=\int{\rm d}{\bf r}{\rm d}{\bf\Omega}[1-\exp(-u_{l,l'}({\bf r},{\bf\Omega}))]$
is the second virial coefficient of a pair of rods of lengths $l,l'$
averaged over their mutual separation ${\bf r}$ and angle ${\bf\Omega}$ \cite{Onsager}.
Finally, the lagrange multiplier $\mu$ is added to free energy to
fix the total monomer concentration $\sum_l l\rho_l=\rho_m$.

The equilibrium length distribution is obtained by minimizing
the free energy with respect to $\rho_l$:
\begin{equation}\label{rhol}
    \rho_l=\rho_0\exp\left(-a l - \sum_{l'}w(l,l')\rho(l')\right)
\end{equation}
where $\rho_0=\exp(-b)/v_0$ and $a=\alpha-\mu$. In the absence of
inter-filament interactions, the distribution $\rho_l^{(0)}$ is
exponential with an average filament length $\langle l\rangle_0 =
1/a = (\rho_m/\rho_0)^{1/2}$ and $r_\sigma = 1$. Note, that because
of monomer conservation the value of $\alpha$ does not affect the
length distribution and only shifts the monomer chemical potential
by a constant.

The leading contributions to the virial coefficient due to
hard-core repulsions are
$w_{\rm HC}(l,l')=(\pi/2) l l' d + \pi(l+l')d^2$
where $d\ll l,l'$ is the filament diameter \cite{GelbartBenShaul}.
As noted by Ben-Shaul and Gelbart \cite{GelbartBenShaul}, the first term
merely shifts the monomer chemical potential by a constant and
has no effect on the distribution.
The next term is equivalent to a reduction in $\rho_0$ and
leads to an increase in the average filament length \cite{GelbartBenShaul}
while the distribution remains exponential with $r_\sigma=1$.

This is no longer the case when linker-mediated inter-filament attractions are introduced
\cite{Borukhov:PNAS}. The new contribution reads
\begin{eqnarray}\label{watt}
    w_{\rm att}(l,l') &=& \frac{\pi}{2}l l' \delta\left(1-\e^{-u_0}\right) \\
    + &2\pi d\delta&\left\{ 2l_1\left[1 + \frac{\e^{-z}-1}{z} \right]
                          + (l_2-l_1)\left[ 1-\e^{-z} \right] \right\}
                          \nonumber
\end{eqnarray}
where $u_0<0$ is the linker mediated short-range attraction per monomer
and $\delta$ is the range of the attraction.
The first term is the contribution from configurations where the filaments
cross at large angles while the second term is the contribution from parallel
configurations
with $z\equiv (l_1/d)u_0$, $l_1\equiv \min(l,l')$, and $l_2\equiv \max(l,l')$.
The latter is divided into a term independent of $l_2$ due to configurations
where the two rods overlap only partially
and configurations where the short filament is fully adjacent to the long filament.

Eqs. \ref{rhol},\ref{watt} allow one to determine the
length distribution $\{\rho_l\}$.
In order to simplify the calculation we look for a perturbative solution
around the noninteracting distribution of the form $\rho_l=\rho_l^{(0)}(1+\delta_l)$
where $\delta_l\ll 1$. To leading order, the solution is then
\begin{equation}\label{deltal}
    \delta_l=\exp\left(-\delta a\,l - \sum_{l'}w(l,l')\rho_{l'}^{(0)} \right)
\end{equation}
where $\delta a$ is determined from the conservation constraint
$\sum_l l\rho_l^{(0)}\delta_l=0$.

\begin{figure}
\includegraphics[width=3.0in]{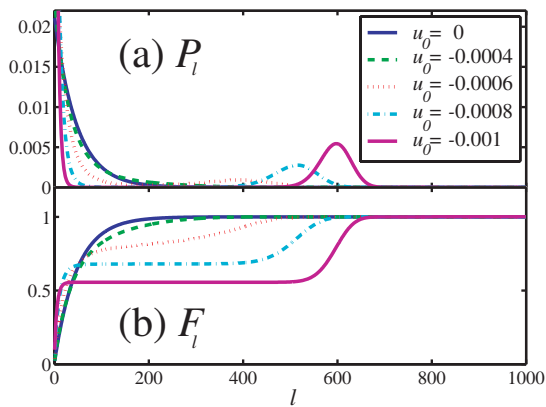}
\vspace{\baselineskip}
\caption{\protect\footnotesize
Length distributions $p_l$ (a) and accumulated distributions $F_l$ (b)
for different values of the attraction strength $|u_0|$ (in units of $k_BT$).
The values used in the calculation are $\rho_m=0.1$, $\rho_0=\exp(-10)$, $\delta=1$, $d=1/2\pi$
and the cutoff value for the filament length is $l_{\rm max}=1000$.}
\label{fig:pl}
\end{figure}

Typical solutions for the normalized length distribution
$p_l=\rho_l/\sum_{l'}\rho_{l'}$ and the accumulated distribution
$F_l\equiv\sum_{l'=1}^l p_{l'}$ for different values of the
attraction strength, $u_0$, are shown in Fig.~\ref{fig:pl}. The
solutions for $u_0 \leq -0.001$ generally agree with the measured
distributions (see inset in Fig.~2) except for the short filament
end, which is beyond the resolution of the experiment. As the
attraction becomes stronger, the distribution develops a peak at
longer and longer filament lengths and with increasing magnitude.
Since $u_0$ is defined as the interaction energy {\em per monomer},
it is the magnitude of $l u_0$ that determines whether the
inter-filament interactions are strong enough to modify the
distribution.

\begin{figure}
\includegraphics[width=3.0in]{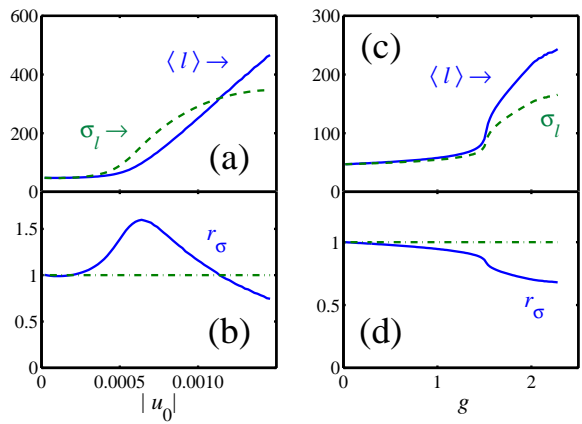}
\vspace{\baselineskip}
\caption{\protect\footnotesize
Average filament length $\langle l\rangle$ and root mean square $\Delta l$ (a)
and their ratio $r_\sigma$ (b) as functions of the attraction strength $|u_0|$
within the perturbative approximation.
The results of the mean field approximation are shown in (c) and (d)
where the horizontal axis is $g\sim \pi u_0^2$.
Same physical values as in Fig.~\ref{fig:pl}}
\label{fig:r}
\end{figure}

The effect on the coefficient of variance $r_\sigma$ is non-monotonic
as shown in Fig.~\ref{fig:r}b.
At first, the ratio increases above one but soon it reaches a maximum and
decreases below one. The initial increase in $r_\sigma$ is due to the bimodal
length distribution at intermediate values of $u_0$,
and the consequence decrease occurs when the longer filaments
dominate the length distribution.

Because of the perturbative nature of the approximation,
it can not correctly describe the behavior in the presence of strong attractions.
In this limit, of strong interactions, the behavior can be studied
within a {\em mean field}-like approximation where the main assumption is that the sum
in the exponent of Eq.~\ref{rhol}
is dominated by the "typical" filament length $l^*=\langle l\rangle$.
It is helpful to expand the attraction term $w(l,l')$ in powers of $u_0$
and neglect a term proportional to $l_1^3$ that is only significant 
in the immediate vicinity of $l^*$.
The length distribution is then of the form
\begin{equation}\label{rholstar}
    \rho_l \simeq \left\{ \begin{array}{l}
       \rho_0 {\rm e}^{-a l + (g l^*\rho^*)l^2} ~ l\le l^* \\
      \rho_0 {\rm e}^{-(a- g {l^*}^2\rho^*)l}  ~ l\ge l^*
    \end{array}\right.
\end{equation}
where $g\sim \pi u_0^2$.
   It is now apparent from Eq.~\ref{rholstar} that while the length distribution of long filaments ($l\ge l^*$) remains exponential, that of short filaments ($l\le l^*$)
shows two local maxima, first at $l=0$ and then at $l=l^*$.
Furthermore, while for small values of $|u_0|$ the distribution reduces to the original
exponential one, at large values of $|u_0|$ the distribution shows narrow peak at $l=l^*$.
With three unknowns $a$,$l^*$,$\rho^*$, the distribution function, $\rho_l$,
can be calculated by solving the following three equations:
(i) the monomer conservation condition $\int{\rm d}l\,l\rho_l=\rho_m$,
(ii) the consistency condition $\rho^*= \rho_0 \exp(-a+ g {l^*}^3\rho^*)$,
and (iii) the condition that $\langle l\rangle=l^*$.
As shown in Fig.~\ref{fig:r}, the results of this calculation (valid only at large values of $|u_0|$) qualitatively confirm the results of the perturbative approach.

\begin{center}
{\bf Conclusions}
\end{center}
In summary, we have shown that in the presence of short-ranged
attractions the coefficient of variance of the length distribution
of F-actin is decreased due to an increasing tendency of the filaments
to align parallel to each other even without chemical
crosslinking. The phenomena is quite general and does not depend
on the exact details of the attraction mechanism: the attractions
can be induced by multivalent counterions or can be electrostatic
by nature; they can occur in a crowded environment like the cell
as a result of depletion mediated attractions; or they can be
induced by linker proteins, which attach actin filaments to each
other in the cytoskeleton and in other filament aggregates.


{\em Acknowledgments:}
We would like to thank Oleg Krichevsky for access to equipment
and Nir Gov for useful discussions.
Partial support from the German-Israel Foundation and the Israeli Science Foundation
is gratefully acknowledged.

\end{document}